\documentstyle[aps]{revtex}

 \def\bar{\overline}          
\def\ua{u^{P}} \def\uv{u^{S}}

\def\fe{{\cal E}}
\def\uo{u^0}

\begin{document}

\begin{flushright}
AMES-HET-98-05 \\
BNL-HET-98/20 \\
\end{flushright}
\bigskip
\begin{center}
{\large\bf Extracting information on CKM phases, electro-weak 
penguins and new physics from $B\to VV$ decays} \\
\vspace*{0.25 in}
{\large\bf Draft 27.0}
David Atwood$^a$
and Amarjit Soni$^b$
\end{center}

\bigskip

\begin{flushleft}
$a$) Dept.\ of Physics and Astronomy, Iowa State University, Ames IA\ \
50011\\
$b$) Theory Group, Brookhaven National Laboratory, Upton, NY\ \ 11973
\end{flushleft}
\bigskip

\begin{quote} {\bf Abstract}:  
We derive constraints for $B\to V_1V_2$ modes ($V_{1,2}={}$vector meson)
that allow a model independent quantitative assessment of the contributions
from electroweak penguins and/or new physics. Interplay of direct CP with
oscillation studies then 
may lead to the extraction of the angle $\alpha$, 
using $B\to K^\ast\omega(\rho)$ and $B\to\rho\omega(\phi)$. 
Any reservation one may have can be explicitly
verified in a  model independent way by assuming only isospin
conservation. We also briefly mention how the method can be used to
extract $\gamma$ via $B_s$ decays: $B_s\to K^\ast\rho$, $\bar K^\ast K^\ast$.

\end{quote}

\vspace*{0.5 in}

\section{Introduction}

Recent evidence from CLEO~\cite{cleoref} indicates that the long sought
after penguin decays occur at the appreciable rate of about $10^{-5}$. In
both $b\to s$ and $b\to d$ transitions interference between the
tree and the strong penguin is expected to
lead to  CP violation effects. Two of the most important applications
of these rare hadronic decays are: 1) determination of the phases of
the unitarity triangle \cite{article}, and 2) test the presence of
non-standard physics. In this paper we will show that among such
$B$-decays those involving two vector mesons in the final state can be
very useful for attaining these goals.

Following Gronau and London \cite{gronautwo}, a general strategy for
extracting CKM angles
from modes that result from the interference
between tree and penguin is to exploit the fact that the strong penguin
and the tree have different isospin transformation properties.
For example, the strong penguin $b\to d$ has $\Delta I=1/2$ whereas the
interfering tree ($b\to u\bar ud$) has both $\Delta I=1/2$ and 3/2.
%
%
%
%
In general, the virtual gluon produced by the penguin contributes a net 
isospin of 0. 
By a suitable choice of combinations of various exclusive final
states
that result from the  
%
%
%
%
quark level transition, 
a separation of the pure $\Delta I=3/2$ piece becomes possible 
yielding the angle $\alpha$~\cite{angles}.

%
%
%
%

If there is a substantial contribution from electro-weak penguins (EWP),
which produce a virtual $Z$-boson or photon, this method fails because the
$Z$-boson and the photon, unlike the gluon, can carry isospin ($I=1$)
component 
yet it has the same weak phase as the strong penguin
\cite{fleischer}.  In the $b\to d$ transition then, both the
tree and the EWP will generate $\Delta I=1/2$ and $\Delta I=3/2$.  Thus in
the standard model, from isospin considerations alone, it is not possible
to isolate EWP from the tree amplitude.
%
%
%
%
%
%
%
%
%
%
%
%
%
%
%
%
%
%

%
%
%

In this paper we suggest what might be the next best thing. We consider
the decay $B\to V_1V_2$ (i.e.\ two vector particles) and, taking advantage
of the information present in the decay distributions of the vector
particles~\cite{valen}, we derive a set of three ratios $\{R_1$, $R_2$,
$R_3\}$ which monitor EWP contamination. In particular if $R_i\neq 1$ then
it is implied that an EWP process (or non-SM physics) makes a significant
contribution.  Decays where EWP are small may thus be identified as
candidates for the extraction of CKM phases (in particular $\alpha$). In 
these cases, as we will show, the value of the CKM phase can
be determined by combining information from direct CP violation with that
from oscillation studies.

%
%
%
%
%
%
%

We focus on four particular examples where this method may be 
applied to determine $\alpha$:
In the case of $b\to s$ transitions:
$K^\ast+\omega$ and $K^\ast+\rho$
while in the case of $b\to d$ transitions: 
$\rho+\omega$ and $\rho+\phi$.  Even
if it turns out that every case has significant 
contamination, important information about the magnitude of these
EWP may still be obtained in these reactions which
should be valuable in its own right. 
For instance, if the contaminating effects
are much larger than anticipated, they may represent 
evidence for
physics beyond the SM.

%
%

The outline of the rest of this paper is as follows. In Section 2 we
describe the crux of our analysis where we determine the phase difference
between each of the meson decays and the corresponding tree amplitude. As
a byproduct of this analysis we obtain a condition which is sensitive to
the effect of EWP amplitudes. In section 3 we describe the extraction of
the necessary information to perform this analysis from experiment.
Then the method for determining the CKM phase $\alpha$ by combining
information from oscillation experiments is discussed.  In
Section 4 we consider various specific examples: $B\to K^*\omega$, 
$K^* \rho$, $\rho\omega$ and $\rho\phi$ and in Section 5 we present
conclusions and briefly mention how the method may also be used to extract
$\gamma$ from $B_s\to K^\ast\rho$ and $\bar K^\ast K^\ast$.

\section{General Formalism}

%
%
%

%
%
We first discuss the general mathematical framework that we 
will use to find the phase between the meson transitions and the quark
level tree diagram.
The cases we will consider here consist of two amplitudes which are
related 
in some way
by isospin, in particular a $B^-$ decay which we denote $u_1^{h}$
and a $\bar B^0$ decay $u_2^{h}$.  We denote the corresponding final
states of the two decays as $f_1$ and $f_2$.  Here, the superscript $h$
indicates the helicity of the vector particles, that is $u_i^h$ is
the amplitude for the decay with the final state
$f_i^h=(V_1^h V_2^h)_i$, $h=-1$, $0$, $+1$. In
addition, one has the conjugate amplitudes $\bar u_1^{h}$ and $\bar
u_2^{h}$ for $B^+$ and $B^0$ decays respectively.

In order to proceed, we need to construct, in the absence of EWP, a
combination of the two amplitudes which receives contribution only from the
tree.  
Such a component can be isolated since only the tree contributes a term to
the effective hamiltonian with $\Delta I=1$ (for $b\to s$) or $\Delta I=3/2$
(for $b\to d$).
%
%
In contrast, for these two transitions, the strong penguin  contributes
to $H_{eff}$ pieces with $\Delta I=0$ and $\Delta I=1/2$ respectively. 
Therefore a combination of the amplitudes which has 
isospin properties as that of the pure tree will contain only the weak
phase of the tree. We 
denote such a combination $c_1 u_1^h+c_2 u_2^h$. This amplitude and its
CP-conjugate will be related as follows:

\begin{equation}
(c_1 u_1^h+ c_2 u_2^h)e^{-i{\delta_T}} =
(c_1 \bar u_1^{-h}+ c_2 \bar u_2^{-h})e^{+i{\delta_T}}
\label{ui_system}
\end{equation}

\noindent 
%
%
where  $\delta_T$ is the weak phase of the tree. Indeed the value of 
$\delta_T$ will depend on the phase convention for particle versus 
antiparticle decays. The physical observables will not depend on this 
convention, though. In practice the same convention 
will also enter the phase of $B\bar B$ oscillation and 
in combination this dependence will cancel.

%
%
%
%
%
%
Constructing such a relation is
straightforward if one of the final particles is an isoscalar. 
More generally, such a relation can be constructed if and only if: 
\begin{enumerate}
\item
 At least two amplitudes related by isospin are involved.
\item
The strong penguin effective Hamiltonian can contribute only one
isospin amplitude to the final state.
\end{enumerate}

\noindent

For example, suppose we have a system of decays
consisting of $n_1$ different states from $B^-$ decays and 
$n_2$ different states from $\bar B^0$ decays where all of the decays
in question 
$B^-\to f_1^i$, $i=1,2\dots n_1$
and 
$\bar B^0\to f_2^i$, 
$i=1,2\dots n_2$ are related by isospin 
(for instance, all the various charge combinations of $K^*\rho$).
If this is a case which satisfies (1) and (2), then there will be only one
strong penguin amplitude, $U_P$, in the above system.
Therefore, one can write the amplitudes for these decays as:

\begin{eqnarray}
u(B^-\to f_1^1) &=& r_1^1 U_P + T_1^1 \\ \nonumber
u(B^-\to f_1^2) &=& r_1^2 U_P + T_1^2 \\ \nonumber
\dots                                 \\ \nonumber
u(\bar B^0\to f_2^1) &=& r_2^1 U_P + T_2^1 \\ \nonumber
u(\bar B^0\to f_2^2) &=& r_2^2 U_P + T_2^2 \\ \nonumber
\dots                               
\end{eqnarray}

\noindent 
where $U_P$ is the strong penguin amplitude, $T_j^i$ are the various tree
contributions to these amplitudes where the subscript ($j=1,2)$ specifies
the initial $B$ state and the superscript ($i=1,2\dots$) designates the final
state. Here $r_j^i$ are coefficients derived
from $SU(2)$ of isospin (i.e.\ Clebsch-Gordon coefficients).  If we now
take any two amplitudes, for instance, $u(B^-\to f_1^1)$ and $u(\bar B^0\to
f_2^1)$, we can write a relation of the type:

\begin{eqnarray}
r_2^1 u(B^-\to f_1^1)
-
r_1^1 u(\bar B^0\to f_2^1)
=r_2^1 T_1^1 - r_1^1 T_2^1
\end{eqnarray}

\noindent
where now the right hand side only contains  tree amplitudes and so has the
weak phase of the tree. This leads to a relation of the form of
eq.~(\ref{ui_system}) where:
 
\begin{eqnarray}
c_1&=&r_2^1 
\nonumber\\
c_2&=&-r_1^1 
\end{eqnarray}

Let us now survey decays of the type $B\to V_1V_2$. In the case of $b\to
s$ transitions the penguin amplitude is $\Delta I=0$ and therefore in all
cases there will be only one penguin amplitude. In particular, this will be
true for $K^*\omega$ and $K^*\rho$. In the latter case there are four
related final states. Since, in principle, we only need two final states
for our analysis, we may, therefore,  choose that pair of final state to
enter into eq.~(\ref{ui_system}) which we expect to be the least effected by
EWP contributions, as will be discussed in section 4. 

In the case of $b\to d$ transitions, the penguin amplitude is $\Delta
I=1/2$. In principle, this can transform the $B$ isodoublet into a $I=0$ or
$I=1$ final state; thus there are two 
possible penguin amplitudes. If $V_1$ is an
isovector and $V_2$ is an isoscalar (e.g.\ $B\to \rho\omega$) then there is
only an $I=1$ final state and a relation of the form eq.~(\ref{ui_system})
may be constructed.  On the other hand $B\to \omega\omega$ does not work
because there is only one amplitude involved while $B\to \rho a_1$ fails
since there are two penguin amplitudes leading to $I=0$ and $I=1$ final
states.

%
%
%

If eq.~(\ref{ui_system}) can be established, then the extraction of
information about phases in the CKM matrix proceeds in three steps. 

\begin{enumerate}

\item First, as discussed in section 3, the study of the angular
distributions of the decay products of the two vector particles in each
reaction will give us the magnitudes of the helicity amplitudes $|u_i^h|$
and the phases between pairs of helicity amplitudes that lead to a common
final state (and thus interfere). 

\item Secondly, as we will describe below, we will use eq.~(\ref{ui_system})
to obtain the phase difference 
between $u_1^h$ and $u_2^h$ (and likewise between $\bar u_1^h$ and $\bar
u_2^h$).  At this stage three conditions allow us to check the consistency
of the assumption that EWP contamination is not significant. 

\item Finally, as
we discuss in section 3, if $u_2$ is the decay amplitude to a self
conjugate mode, an oscillation experiment fixes the phase between $u_2$
and $\bar u_2$ so that now the phase between all pairs of amplitudes becomes
known and information about $\delta_T$ 
(in combination with the $B\bar B$ oscillation phase)
may then be recovered.

\end{enumerate}

In order to carry out the third step of this program, the neutral $B$
decay must be to a self conjugate state. In examples such as $B\to
\phi\rho$ this requirement is  met since the state $\phi\rho^0$
is self conjugate. On the other hand, when one of
the final state 
particles is a $K^{0*}$, for example $B\to K^*\omega$, this requirement
is only met for the $K^*$ decays into a CP eigenstate, that is $K^{0*}\to
K_S \pi^0$.

%
%

In the case of $B\to K^* \rho$ we can only perform an oscillation
experiment on the final state $\bar B^0\to \bar K^{*0}\rho^0$.  We will
argue, however, that the amplitude for $\bar B^0\to K^{*-}
\rho^+$ is more likely to be free of EWP effects. So, in section 4, we
will invoke additional isospin relations based on the assumption that
the EW Hamiltonian has no $\Delta I=2$ piece in order to interpret the
oscillation experiment for $\bar B^0\to \bar K^{*0}\rho^0$ to obtain the
desired weak phase of the quark level transition. This more complicated
strategy is necessitated by the realization that the decay $\bar B^0\to
\bar K^{*0}\rho^0$ is more susceptible to EWP 
contamination than modes containing
the charged $\rho$. Should this not be the
case then we may proceed more directly applying our method to this final
state (i.e.\ $\bar K^{\ast0}\rho^0$) together with any one of the other
$K^\ast\rho$ modes.


%
%
%
%
Let us now discuss how to use equation eq.~(\ref{ui_system})  in order to
extract the phases between the helicity amplitudes $u_1^h$ and $u_2^h$. 
To start with, for each specific final state the angular distributions of
the vector
decays will give the magnitudes $|u_i^h|$ as well as the relative phases
between two amplitudes of differing helicities for the same final state
(i.e. the phase between $u_i^{h_1}$ and $u_i^{h_2}$).
One may obtain this information by fitting the experimental data to the
distribution given in the next section (see eq.~(\ref{eqseven})).  We will 
denote the
relative phases between the two helicity states ($h_1$ and $h_2$) of the same
final state ($f_i$ for $i=1,2$)  by $\phi_i(h_1,h_2)=arg(u_i^{h_1}
u_i^{h_2*})$; $\bar \phi_i(h_1,h_2)=arg(\bar u _i^{h_1} \bar u_i^{h_2*})$. 
This information together with eq.~(\ref{ui_system}) gives us the system
of equations we must solve for the relative phases of the amplitudes. 

%
%
%
%
%
%
%
%
%
%
%
%
%
%
%

Before proceeding to solve this system, it is useful to factor out the
tree weak phase $\delta_T$ and rewrite the equations in 
terms of the quantities $v_i^h$.  
We thus define  $v_i=e^{-i\delta_T}u_i$; $\bar
v_i=e^{+i\delta_T}\bar u_i$. The system then becomes

\begin{equation}
(c_1 v_1^h+ c_2 v_2^h) =
(c_1 \bar v_1^{-h}+ c_2 \bar v_2^{-h})
\label{vi_system}
\end{equation}

\noindent where $|v^h_i|=|u^h_i|$ and $\phi_i(h_1,h_2)=arg(v_i^{h_1}
v_i^{h_2*})$ (and likewise for the conjugate case) are
quantities that may be determined experimentally.

%
%
%

It is convenient to express the above in terms of parity eigenstates,
which we denote $v^k$, where $k=0$, $P$ or $S$.  This basis is defined as
$v^S= (v^{+1}+v^{-1})/\sqrt{2}$, $v^P= (v^{+1}-v^{-1})/\sqrt{2}$ and $v^0$
is common to both bases. The system of equations~(\ref{vi_system}) thus
becomes: 

\begin{eqnarray}
(c_1 v_1^{0,S,P}+ c_2 v_2^{0,S,P})
&=&\pm(c_1 \bar v_1^{0,S,P}+ c_2 \bar 
v_2^{0,S,P})
\label{vi_system2}
\end{eqnarray}

\noindent
where $\pm=+$ for the $0$ and $S$ cases and $\pm=-$ for  the $P$
case. 

%
%
%
%

Experimental data gives us 
the phase between $v_i^{k_1}$ and $v_i^{k_2}$ for a given final state
$f_i$ where $k_1,k_2\in\{0,P,S\}$ and likewise for $\bar v$.  Thus, all we
need to know is the phase of $v_1^0$, $v_2^0$, 
$\bar v_1^0$ and $\bar v_2^0$ to fix
all of the phases of $v_i$.

Let us denote 
these phases for the $0$-helicity amplitudes by
$\psi_i=arg(v_i^0)$ and $\bar \psi_i=arg(\bar v_i^0)$
(where $i=1,2$).
Clearly,
then the system of equations~(\ref{vi_system2}) becomes a series
of linear conditions on $\{ e^{i\psi_i}$, $e^{i\bar \psi_i} \}$. 
The solution is given by considering the determinant:

\begin{eqnarray}
\Delta= \left |
\begin{array}{cccc}
x_1&x_2&\bar x_1&\bar x_2         \\
 c_1 |v_1^0|&  c_2 |v_2^0|& -c_1 |\bar v_1^0|& -c_2 |\bar v_2^0| \\
 c_1|     v_1^{S}|e^{i     \phi_1(S,0)}&
 c_2|     v_2^{S}|e^{i     \phi_2(S,0)}&
-c_1|\bar v_1^{S}|e^{i\bar \phi_1(S,0)}&
-c_2|\bar v_2^{S}|e^{i\bar \phi_2(S,0)}| \\
 c_1|     v_1^{P}|e^{i     \phi_1(P,0)}&
 c_2|     v_2^{P}|e^{i     \phi_2(P,0)}&
 c_1|\bar v_1^{P}|e^{i\bar \phi_1(P,0)}&
 c_2|\bar v_2^{P}|e^{i\bar \phi_2(P,0)} 
\end{array}
\right | \label{det1}
\end{eqnarray}

%
%
%
%

\noindent where a solution exists if and only if 
the three ratios

%
%

\begin{eqnarray}
R_1&=&
\left | {\partial \Delta / \partial  
x_1 
} \right | 
/
\left | {\partial \Delta / \partial  
\bar x_1 
} \right | 
\nonumber\\ 
R_2&=&
\left | {\partial \Delta / \partial  
x_2 
} \right | 
/
\left | {\partial \Delta / \partial  
\bar x_2 
} \right | 
\nonumber\\
R_3&=&
\left | {\partial \Delta / \partial  
x_1 
} \right | 
/
\left | {\partial \Delta / \partial  
\bar x_2 
} \right | 
\label{Rrelation}
\end{eqnarray}

\noindent
satisfy

\begin{eqnarray}
R_1=R_2=R_3=1
\label{condition2}
\end{eqnarray}

\noindent
If this condition holds, 
the required phases are then

\begin{eqnarray}
\psi_i=\xi_0+arg\left(\partial \Delta / \partial      x_i \right);
\ \ \
\bar\psi_i=\xi_0+arg\left(\partial \Delta / \partial  \bar x_i 
\right) 
\label{solution}
\end{eqnarray}

%
%

\noindent where $\xi_0$ is an overall strong phase which cannot be
determined (and does not enter into any of the physics discussed here). 
Clearly  eq.~(\ref{condition2}) may also be
regarded as a test for the presence of EWP or new physics
effects. If there is a contribution from new physics or EWP, the set of three
equations (i.e.\ eq.~(\ref{condition2})) implies that for each helicity
the new contribution satisfies one equation. Unless the new
contribution has the same weak phase and also the same isospin
transformation properties as the tree, this is rather improbable. Thus,
eq.~(\ref{condition2}) provides a good test for the presence of EWP
and/or new physics. We must also emphasize that this test of EWP is
completely model independent since  
%
%
%
it only assumes isospin conservation.

%
%
%
%

The phases of $v^k_i$ are physically meaningful modulo the overall strong
phase $\xi_0$ above. For instance, in the case of $B\to K^\ast\omega$, 
we can interpret 
$arg(v^k_1)$  
and
$arg(v^k_2)$  
as the phases between
the quark level  
$b\to s \bar u u $ 
transition and the meson decays 
$B^-\to K^{*-}\omega$ 
and
$\bar B^0\to \bar K^{*0}\omega$ respectively 
in the helicity combination indicated by ($k$). 
Likewise 
$arg(\bar v^k_1)$  
and
$arg(\bar v^k_2)$  
are the phases 
between
$\bar b\to \bar s \bar u u $ 
and
$B^+\to K^{*+}\omega$ and
$B^0\to K^{*0}\omega$.

%
%
%
%
%
%

Although at this point we know the phases of all the $v_i^k$ and $\bar
v_i^k$ ($i=1,2$, $k\in\{0,P,S\}$) we still do not know the phase
differences between $u_i^k$ and $\bar u_i^k$ since we do not know
$\delta_T$. Indeed in the context of the standard model it is 
$\delta_T$ 
which we wish to know since it is derived from the CKM matrix.  
Since $\delta_T$
cannot be obtained from the experimental information which we have
included so far, one needs some additional data which
depends, in particular, on the phase difference between a particle and
anti-particle decay.

In the examples we consider, the decay from the neutral $B$-meson provides
an opportunity to do this, since in that instance, oscillation effects
allow the interference of $B^0$ and $\bar B^0$ decay amplitudes. Thus, we
are able to interfere the amplitudes $u_2^k$ and $\bar u_2^k$ if the final
state $f_2$ is a CP eigenstate, for instance, $B^0$, $\bar B^0\to \phi\rho^0$. 
In the following section we show that from observing the decay of $B^0$ and 
$\bar B^0$ as a function of time, it is possible to extract the quantity
$\sin(\zeta-2(\hat\beta+\delta_T))$ where the angle $\zeta$ is a function
only
of $v_i^k$ which can be determined as described above and $\hat \beta$ is
the phase
from the CKM matrix inherent in neutral $B$ oscillations 
(using the same convention in which $\delta_T$ is defined in). 
If $\zeta$ is determined then the quark-level quantity $\hat\beta+\delta_T$ 
($\hat\beta+\delta_T=\beta+\gamma=\pi-\alpha$ in the examples we consider
in the standard convention of \cite{wolfmatrix}), which 
depends only on the CKM matrix~\cite{angles}, may thus be extracted up to
the ambiguity  
of the sine function. Specifically, if we adopt the CKM phase convention
of~\cite{wolfmatrix}, then in the standard model $\hat\beta=\beta$ and
$\delta_T=\gamma$.

In summary, the extraction of $\hat\beta+\delta_T$ proceeds as follows. 
First we must determine from the angular distributions of the decays the
magnitudes of $|v_i^{(0,P,S)}|$ and $|\bar v_i^{(0,P,S)}|$ as well as the
phases $\phi_i(V,0)$, $\phi_i(A,0)$, $\bar\phi_i(V,0)$ and
$\bar\phi_i(A,0)$.  We then check that there is no EWP contamination via
eq.~(\ref{condition2}). If this is satisfied, then for any value of $i$,
we can obtain the phases of $v_i$ and $\bar v_i$ from eq.~(\ref{solution}) from
which we can calculate $\zeta$ as we will describe below. Observing
oscillation effects in the neutral $B$ decay will then allow us to obtain
$\hat\beta+\delta_T$ which yields the phase $\alpha$ of the unitarity
triangle.

\section{Experimental Procedure}

Let us now discuss the experimental observables which are needed to
perform the above analysis. The basic ingredient will be the study of
correlations between the decay distributions of the two vector mesons or,
equivalently, the correlation of their polarizations.

First, let us consider the
case where the vector meson $V$ decays to two pseudo-scalars $V\to
P_1P_2$; for instance, $\rho\to\pi\pi$, $\phi\to K\bar K$ and $K^*\to
K\pi$. Then the polarization vector $\fe_V$ in the rest frame of $V$ can
be taken to be parallel to the momentum of one of the pseudoscalars, $\vec
\fe_V \propto \vec P_{P_1}$.  We are not concerned about the sign of
$\vec \fe_V$ since it will not enter into the analysis below. 
The case of $\omega$ decaying to $3\pi$ is similarly self analyzing since
if  $\omega\to \pi^+\pi^-\pi^0$, 
the polarization is related to the momenta of the pions by
$\vec \fe_\omega\propto (\vec
p_{\pi^+}\times \vec p_{\pi^-})$, in the rest frame of the $\omega$.

In the $V_i$ rest frame, denote the angle between $\vec \fe_i$ and $-\vec
P_B$ (the three momentum of the $B$ meson) by $\theta_i$. Let us define
$\Phi$ to be the azimuthal angle from $\vec \fe_1$ to $\vec \fe_2$ in the 
rest frame of the $B$ about $\vec P_{V_1}$ such that $\sin\Phi\propto
(\vec \fe_1\times\vec P_{V_1})\cdot \vec \fe_2$. If we define $y_i=\sin\theta_i$
and $z_i=\cos\theta_i$ then the angular distribution of the decays in terms
of $\{\theta_1$, $\theta_2$, $\Phi\}$ is: 

\begin{eqnarray}
{d^3\Gamma/(dz_1 dz_2 d\Phi)} &=& |\uo|^2 z_1^2 z_2^2
+  y_1^2 y_2^2 (|\uv|^2\cos^2\Phi+|\ua|^2\sin^2\Phi) \nonumber\\
&& +2 Re(\uo(\uv)^*) y_1 y_2 z_1 z_2 \cos\Phi   \nonumber\\
&& -2 Im((\uo)^*\ua) y_1 y_2 z_1 z_2 \sin\Phi   \nonumber\\
&& +2 Im(\uv(\ua)^*) y_1^2 y_2^2 \sin\Phi\cos\Phi \label{eqseven}
\end{eqnarray}

From an experimental study of the distribution of the decays, one can
extract the  quantities $|u^0|$, $|u^S|$, $|u^P|$ as well as
$\cos\phi(0,S)$, $\sin\phi(P,0)$ and $\sin\phi(S,P)$; the latter three
correspond to interference terms of the type $u^{(0)}u^{(S)\ast}$. Note that
there is a two fold ambiguity in the determination of the actual phase
differences since either $\{\phi(0,S)$, $\phi(P,0)$, $\phi(S,P)\}$ or
$\{-\phi(0,S)$, $\pi-\phi(P,0)$, $\pi-\phi(S,P)\}$ will explain a given set
of data. When the data for all the helicities and for the decays of the
neutral and charged $B$ to specific modes is considered together, however,
eq.~(\ref{condition2}) 
should only work for one of the two cases.  Note
also that these phase angles satisfy the condition,   
$\phi(0,V)+\phi(A,0)+\phi(V,A)\equiv 0\ {\rm mod}\ 2\pi$, which is a useful
constraint on interpreting the experimental data.

%
%
%
%


In the above distribution, CP violation will be manifest by the difference
between $d^3\Gamma(z_1,z_2,\Phi)$ for a $\bar B^0$ or $B^-$ meson and
$d^3\Gamma(z_1,z_2,-\Phi)$ for the conjugate meson decay.  The two
manifestly $P$-odd interference terms $\propto\ua$ represent CP violating
effects which are P-odd C-even.  Further, these terms are odd under
``naive time reversal ($T_N$)'', defined as the inversion of momenta and
spins without the interchange of initial and final states required under
$T$.  Such effects are present even if there are no absorptive phases.  In
contrast, the other four CP violating terms are even under $T_N$ and so
only present if there are absorptive phases.  Two possible sources for
such phases are (1)~the result of rescattering at short
distances~\cite{pertphase} or (2)~at long distances~\cite{donoghue,mancp}.

%
%

%
%
%
%

Let us now discuss the problem of extracting the CKM phase 
$(\hat \beta + \delta_T)$
through the observation of oscillations effects in the decay of neutral
$B$ mesons assuming that, through the use of eq.~(\ref{condition2}), it has
been demonstrated that EWP contributions are negligible.  As indicated above, we assume
that $u_2$ represents the decay from the neutral $B$-meson, i.e.\ $\bar
B^0\to V_1V_2$  while $\bar u_2$ the decay $B^0\to \bar V_1\bar V_2$.
In such an oscillation experiment, we will assume that at a point in time,
which we define to be $t=0$, the flavor of the neutral $B$ meson is known
to be either $\bar B^0$ or $B^0$ due to some tagging event. At an $e^+e^-$
machine, sitting at the $\Upsilon(4S)$, this tagging event 
would be the decay of the associated meson to a
final state of unambiguous flavor. For instance, if the partner decayed to
$e^+\nu_e D^-$ at $t=0$ then the meson in question must be a $\bar B^0$ at
$t=0$. At a hadron collider, the tagging event would generally occur at
the moment of creation. For example, if $p+p\to B^+\bar B^0+X$ at $t=0$
then the flavor of the neutral $B$ meson is unambiguously fixed at that
point in time. In the following, therefore, negative values of $t$ are
allowed in $e^+e^-$ experiments while only positive values will apply to
hadronic collisions.

Below we will consider only the total decay rate as a function of time $t$. 
The generalization to decay distributions as a function of time is 
straightforward but it is probably much more difficult experimentally to 
use such information. In any case the extraction of phases of the CKM 
angle can be made from the inclusive time dependent rate.

Let us denote $\Gamma_B$ to be the total width of the neutral $B$ meson
and $g(t)=d \Gamma(\bar B^0(t)\to V_1V_2)/dt $ to mean the differential
rate that a meson, identified as a $\bar B^0$ at $t=0$, decays to $V_1V_2$
at time $t$. Likewise we denote $\bar g(t)=d\Gamma(B^0(t)\to \bar V_1\bar
V_2)/dt$. At $t=0$ let us define $\hat g \Gamma_B= r g(0)$ and $\bar{\hat
g} \Gamma_B= r \bar g(0)$, where, $r=1$ in cases when only $t\geq 0$ (i.e.
hadronic colliders) is allowed and $r=2$ when both signs of $t$ are
present (i.e. $e^+e^-$ colliders). Here $\hat g$ and $\bar{\hat g}$ are the
decay rates that would be present in the absence of oscillations.  These
may also be obtained in self-tagging situations which apply to some cases
as discussed below.

Clearly, interference is only possible if the states $V_1V_2$ 
and $\bar V_1\bar V_2$ eventually cascade down to the same final state. 
The simplest situation where this applies, and the case we shall consider 
here is when $V_1V_2$ is an eigenstate  of $C$ (charge conjugation) 
with eigenvalue $\lambda=\pm 1$.

%
%
From the analysis of section 2 we 
know the phase of each of the meson decay amplitudes with respect to the
tree graph. Using this information we can now write the following expression 
for the total decay rate to the final state under consideration as a function
of time:  
%
%

\begin{eqnarray}
(g(t)+\bar g(t))/2 &=& (\hat g +\bar{\hat g})
\Gamma_B e^{-\Gamma_B|t|}/2 \nonumber\\
(g(t)-\bar g(t))/2 &=& (\hat g +\bar{\hat g})
\Gamma_B e^{-\Gamma_B|t|} (T_c\cos(\Delta m\  t)+T_s\sin(\Delta m\  t))/2
\nonumber\\
\ 
\end{eqnarray}

\noindent where $\Delta m$ is the $B_0\bar B_0$ mass difference and

\begin{eqnarray}
T_c= \sum_k (|v^k_2|^2-|\bar v^k_2|^2)/{\cal V};\  \ \ \ 
T_s= -\lambda Im \left [ e^{-2(\hat\beta+\delta_T)} \sum_k\bar v^{k}_2 
v^{k*}_2/{\cal V}\right ] 
\end{eqnarray}

\noindent where, again, $k$ can be taken to be $0$, $S$, $P$ with ${\cal
V}=\sum_k (|v^k_2|^2+|\bar v^k_2|^2)$. 
If we denote $R \exp ({i \zeta})  =\sum_k\bar v^{k}_2 v^{k^\ast}_2$ 
then the above expression for $T_s$ may be re-written:

\begin{equation}
T_s=-\lambda(R/{\cal V})\sin\left( \zeta - 2(\hat\beta+\delta_T) \right )
\end{equation}

\noindent where the values of $R$ and $\zeta$ may be obtained  once
the phases of $v_i^k$ are determined from eq.~(\ref{solution}). Thus, from
the experimental determination of $T_s$, one obtains, up to a four fold
ambiguity, the value of $\hat\beta+\delta_T$. The additional solutions
which
produce identical results to a given value of $\hat\beta+\delta_T$ are
$\{\pi+\hat\beta+\delta_T$, $\pm \pi/2+\zeta-\hat\beta-\delta_T\}$. 
%
%
%
The latter two possible solutions which involve $\zeta$
could be eliminated if a different mode with a different value of $\zeta$ 
were considered. 
The other spurious solution  requires that the quadrant of 
$\hat \beta+\delta_T$ be separately known and 
cannot be eliminated via this
kind of oscillation experiment since the angle enters as 
$2(\hat \beta+\delta_T)$.

\section{Specific Examples}

Consider now the application of this method to a few cases relevant to
$b\to s$ and $b\to d$ penguin transitions. The first example is $B\to
\omega K^*$. Here the underlying process is a $b\to s u\bar u$ or $b\to s
d\bar d$ transition. The strong penguin is $\Delta I=0$ and the tree $b\to
s u\bar u$ has both $\Delta I=0$, $1$.  
Define 
$u_1={\cal M}(B^-\to K^{*-}\omega)$ 
and 
$u_2={\cal M}(\bar B\to \bar K^{*0}\omega)$.
%
%
%
From isospin considerations we obtain
$c_1=-c_2=1$ since $(u_1-u_2)$ is proportional to the $\Delta I=1$ 
amplitude only and so must have only the weak phase of the tree.
%
%
%
In this case $\delta_T=\gamma$, so
that $\hat\beta+\delta_T=\beta+\gamma =\pi-\alpha$, in the above. Thus, if 
the contamination
of EWP is small, the angle $\alpha$ may be extracted following the
procedure outlined above.  The degree to which such contamination is
present may be gauged by checking the condition in eq.~(\ref{condition2}).

One feature of the neutral $B$ meson in this case is that one may control
whether oscillation effects are present or not by selecting the decay mode
of the $K^{0*}$; thus if $\bar B^0\to \omega \bar K^{*0}\left [\to
K_s\pi^0\right ]$ the final state is an eigenstate of $C$ so this mode may
be used to extract $T_s$. If, on the other hand, $\bar B^0\to \omega \bar
K^{*0} \left [\to K^-\pi^+\right ]$ then clearly the flavor of the initial
state is determined from the final state and oscillation effects are
absent allowing the direct determination of $|v_2^k|$. Unfortunately, it is
not quite clear that the EWP are small; some estimates\cite{fleischer} of
color allowed EWP to such final states indicate that the contamination 
may be $O(10\%)$.

%
%
%
%
%

It may, however be possible to select final meson states where
EWP effects are likely to be small based on 
the assumption that color suppression tends to render them unimportant. 
With that in mind, observe that the 
contribution to $K^\ast\omega$ by the EWP is
color allowed when both the quarks that result from the virtual $Z$ or
$\gamma$ form the $\omega$. However, note that for this unsuppressed
contribution, the EWP has $\Delta I=0$ since the $Z$ or $\gamma$ are
then converting to an $I=0$ object (i.e\ $\omega$).
The failure of the condition  in eq.~(\ref{condition2}) and
the problem of extracting $\alpha$ which these diagrams cause comes only 
from their $\Delta I=1$ component. 
Therefore this manifestation of the 
color-allowed EWP does not effect the determination of $\alpha$ 
given that (as is the case in the SM)  
the electro-weak and strong penguins have the same weak phase.
The effects which result from 
the $\Delta I=1$ component arise from hadronization where 
one of the quarks from the $Z$, $\gamma$ goes with the $\omega$ and the 
other with the $K^*$. Such diagrams are color suppressed and so their 
contamination on the ability to determine 
$\alpha$ are expected to be only
$O(1\%)$.  However, since our understanding of color suppression is not
reliable it would 
be very  useful to quantitatively ascertain the
EWP through the use of eq.~(\ref{condition2}). Thus, based on all that
we know so far it seems very likely that $K^\ast\omega$ would be a very
good mode for the extraction of $\alpha$. The mild reservation
regarding the presence of EWP can and should be verified through
eq.~(\ref{condition2}).

Another example where color suppression may  reduce the effect of EWP is
in the class of decays $B\to K^*\rho$.  First, consider the case when there
were no EWP. Then each helicity combination,
$0$, $S$ and $P$ behaves like the analogous $K\pi$ system which is
discussed  in~\cite{mancp,gronau}. Furthermore, the cases where EWP would
be color suppressed are those which contain $\rho^\pm$. Thus, if we denote,
$u_1={\cal M}(B^-\to K^{*-}\rho^0)$, $u_2={\cal M}(B^-\to \bar
K^{*0}\rho^-)$, $u_3={\cal M}(\bar B^0\to K^{*-}\rho^+)$ and $u_4={\cal
M}(\bar B^0\to \bar K^{*0}\rho^0)$, the assumption that electro-weak
penguin diagrams are color suppressed and are negligible is equivalent to
saying that $u_2$ and $u_3$ are free of 
EWP.


In this case the application of isospin is somewhat more complicated than 
in the previous case where one of the final state mesons was an 
iso-singlet. We can, however construct a relationship of the desired form 
between the two amplitudes, $u_2$ and $u_3$, by noting that if only $\Delta 
I=0$ contributions were present, $u_2+u_3=0$. This means that more generally
$u_2+u_3$ is proportional to the $\Delta I=1$ transition amplitude and 
will have the weak phase of the tree graph 
although it will be a combination of the amplitude going to a $I=1/2$ and 
$I=3/2$ final state.
It thus follows that

\begin{equation}
v_2^{h}+v_3^{h}=\bar v_2^{\,-h}+\bar v_3^{\,-h}\label{eqeleven}
\end{equation}

%
%
%
%
%
%

\noindent which is a relation of the form
of eq.~(\ref{vi_system}), and so one obtains all the results that follow
from it; in particular, the test for EWP contamination 
eq.~(\ref{condition2}) and the determination of the phases of the amplitudes
eq.~(\ref{solution}). The phases of $v_{2,3}$ and $\bar v_{2,3}$ may 
thus be determined.

If EWP contamination were absent from all of $B\to K^*\rho$ amplitudes then 
one would also have a similar expression 
involving $v_1$ and $v_4$ and could obtain these phases in the same way. 
In particular, we need to know the phase of $v_4$
to obtain $\alpha$
through  an oscillation experiment since $\bar K^{\ast 0}\rho^0$ is the only
case where the final state may be a CP eigenstate (if the neutral $K^*$ 
decays to $K_s$, for instance).
Fortunately, we may obtain an isospin relation which gives us the 
required phases 
in terms of $v_2$ and $v_3$
independent of the possibility of 
EWP contamination since the magnitudes 
$|     v^h_{1,4}|$  and $|\bar v^{-h}_{1,4}|$ are known  
and 
even in the presence of EWP contamination, $v_1$ and $v_4$ are related to 
$v_2$ and $v_3$ through: 

\begin{eqnarray}
v_4^h-v_1^h=(v_2^h-v_3^h)/\sqrt{2};\ \ \ \ \ 
\bar v_4^{-h}-\bar v_1^{-h}=(\bar v_2^{-h}-\bar v_3^{-h})/\sqrt{2}
\label{eqtwelve} 
\end{eqnarray}
%
%

%
%
%
\noindent 
These relations follow since the left and the right side of each
is proportional to the $\Delta I=0$ component of the transition, 
 assuming
that there is no $\Delta I=2$ transition. This latter assumption
would be valid in the SM (to the extent that isospin is conserved
and we are working up to the lowest order in weak interactions) and in most
of its extensions. It follows that in the SM 
these apply even with an arbitrary amount of  electro-weak
penguin contamination to $v_1$ and $v_4$.  
%
%
%
Note also that
unlike~\cite{mancp,gronau}, the measured phases  between the helicity
amplitudes are essential to fix the phases of the amplitudes $v^h_2$ and
$v^h_3$ because with the EWP contamination to $v_{1,4}$ only the relation
(\ref{eqeleven}) between $v_2$ and $v_3$ can be formed. 
The relation (\ref{eqeleven}) between  $v_2$ and $v_3$ 
forms the basis for finding their relative phase and 
eq.~(\ref{eqtwelve}) allows us then to find the relative phases of the
$v_{1,4}$ 
independent of the EWP contamination to 
these amplitudes.

%
%
%
%
%
Since the magnitude and phase of the right hand side of each of
these eq.~(\ref{eqtwelve}) is known, one can solve for the phases
of $v_{1,4}^h$ and $\bar v_{1,4}^{-h}$ up to a 2-fold ambiguity.
%
%
%
%
%
The observed relative phases between the various helicities
in the $v_1$ and $v_4$ channels eliminates this ambiguity.  We can
then use the phase differences between $v_4^h$ and $\bar v_4^h$
and find $\alpha$ from the oscillation data for $\bar B^0\to \bar
K^{0*}\rho^0$ as previously described. 
Once again we stress that
for this analysis for $\alpha$, through $B\to K^\ast\rho$ modes,
to work,        
we must assume only that
EWP are small in the 
color suppressed 
instances, which are 
states containing $\rho^\pm$ (i.e. $u_2$, $u_3$).
This is clearly highly plausible, but in any case is
verifiable  through eq.~(\ref{condition2}). No 
corresponding assumption
regarding $\rho^0$ modes ($u_1$ and $u_4$) is required. 

In the case where the EWP are negligible, it is interesting to compare the 
information that may be learned from the $B\to VV$ decays where there are 
three helicity amplitudes with that from $B\to PP$ and $B\to VP$ decays where
there  is only one.  Cases of the latter type would include $B\to 
K\eta(\eta^\prime)$ or $B\to K\omega$, for instance. 

For $B\to K\omega$, we define $u_1={\cal M}(B^-\to K^{-}\omega)$ and
$u_2={\cal M}(\bar B\to \bar K^{0}\omega)$ but here there is no helicity
dependence and all angular distributions are isotropic so one only knows
the magnitudes of the amplitudes but not their phases. In the absence of
EWP,
$v_1-v_2=\bar v_1-\bar v_2$, which still leaves
free one degree (aside from an overall strong phase) of freedom, the magnitude
$f=|v_1-v_2|$.  

We can, however, infer some inequalities which will apply in these cases. 
The equation among the complex amplitudes: $v_1-v_2=\bar v_1-\bar v_2$ 
implies that:

\begin{equation}
|v_1|\leq |v_2|+|\bar v_1|+|\bar v_2| \label{eqvees}
\end{equation}

\noindent where the four amplitudes may be permuted. 
If these inequalities are not satisfied, then it would mean that there is 
significant 
 contamination from EWP or from some 
source of new physics.

In decays to scalars 
with more complicated structure, for instance $B\to
K\pi$, it is also possible to detect the presence of 
EWP
using
the equation~\cite{mancp}:

\begin{equation}
2|m_1|^2-|m_2|^2-|m_3|^2+2|m_4|^2 =
2|\bar m_1|^2-|\bar m_2|^2-|\bar m_3|^2+2|\bar m_4|^2.
\label{kpirel}
\end{equation}

\noindent where $m_1$ etc. are the amplitudes of the four $K \pi$
modes~\cite{mancp}. If with this relation the EWP
are  confirmed to be negligible, it will
%
%
then be possible to extract $\alpha$ as described
in~\cite{gronau,mancp}.

Returning to the case of $B\to K^*\rho$, we can construct similar 
identities for each helicity:

\begin{equation}
2|u_1^h|^2-|u_2^h|^2-|u_3^h|^2+2|u_4^h|^2 =
2|\bar u_1^{-h}|^2-|\bar u_2^{-h}|^2-|\bar u_3^{-h}|^2+2|\bar u_4^{-h}|^2.
\label{kstrhorel}
\end{equation}

\noindent Again, it is worth noting that these relations
[eqs.~(\ref{eqvees})--(\ref{kstrhorel})] are completely model
independent, they assume only isospin conservation.
It may, for instance, be of particular interest to consider the
case involving $u^{+1}$ and $\bar u^{-1}$ since these would require the
final state $s$-quark to be right-handed and so may be suppressed in the
SM\null. On the other hand, effects from new physics which couple
to right-handed fermions may be enhanced in this channel and so in that
case eq.~(\ref{kstrhorel}) may be sensitive to such contributions.

%
%
%
Note that eq.~(\ref{kstrhorel}) also applies if $h$ represents one of
the parity eigenstates $0$, $V$ or $A$. If one of 
these cases has  a small 
%
%
EWP contribution, $\alpha$ may be
extracted from that case also using the same method as in the $B\to K \pi$
system (with a sign change for barred amplitudes in the $A$ case).

We can also consider the analogous case where there is a $b\to d$ transition,
for example,  $B\to \rho\omega$. Now, the strong penguin  is $\Delta I=1/2$
and the tree process $(b\to d u\bar u)$ contains both $\Delta I=1/2$ and
$\Delta I=3/2$ components. Likewise, possible electro-weak penguin processes
are  $\Delta I=1/2$, $I=3/2$.

Here again $\delta_T=\gamma$ so that the CKM phase we may hope to 
recover through our method is still  $\alpha$. If we define $u_1={\cal M}(B^-\to\rho^-\omega)$
and  $u_2={\cal M}(\bar B \to \rho^0\omega)$, isospin gives
$c_1=-c_2=1$.  Now, color-allowed EWP contribution (i.e.\ $Z\to
\omega$) will not cause any problem as for them $\Delta I=1/2$ as in
the case of the strong penguin. However, the other color-allowed EWP (i.e.\
$Z\to \rho^0$) will be problematic. Thus $B\to \rho\omega$ can only
become a viable method for extracting $\alpha$, if it can be shown,
through eq.~(\ref{condition2}), that EWP contamination is small.

%
%
%

It has been suggested~\cite{donoghue,mancp} that rescattering
effects in exclusive states of the type that we are considering
may be large due to the presence of many intermediate states which
rescatter to such a final state.  If this is true, the quark
content in the final state may differ from that initially present
in the weak decay.  For instance, the tree level transition $b\to
d u\bar u$ could lead to decays like $B\to \rho \phi$ on the meson
level. 
%
%
%
%
%
Here, the
EWP contamination from $b\to d s\bar s$ will, again, not be a
problem since it has the same isospin properties as the strong
penguin. The contamination that might cause a problem will come
from rescattering of the EWP modes $b\to d u\bar u$ and $b\to d
d\bar d$ to $b\to ds\bar s$. This is expected to be extremely tiny
as it originates from Zweig suppressed conversion of the EWP
amplitude.

%
%

In this example the strong penguin and the $b\to d s\bar s$ EWP are
$\Delta I=1/2$ while the tree process has $\Delta I=1/2$ and $\Delta
I=3/2$.  If we define $u_1={\cal M}(B^-\to\rho^-\phi)$ and $u_2={\cal
M}(\bar B \to \rho^0\phi)$, the isospin structure is clearly the same as
$\rho\omega$; so $c_1=-c_2=1$. Again $\delta_T=\gamma$ so that the
analysis will give $\alpha$. Thus final state rescattering of tree
amplitudes in exclusive channels has a nice application here as it leads
to a clean method for obtaining $\alpha$.

\section{Conclusions}

To summarize, in this work we have provided a systematic, model independent
technique for
quantitatively assessing the importance of electro-weak penguins and/or
new physics by studying $B$ decays to two vector particles resulting
from penguin and tree interferences. Our tests only assume isospin conservation.
The modes that do not exhibit such
effects can then be used for extracting the angles of the unitarity
triangle. $B\to K^\ast\omega(\rho)$, $\rho\omega(\phi)$ can all be used
for extracting $\alpha$.

We close with the following two brief remarks:

\begin{enumerate}

\item Our first comment concerns the expected branching ratio. In this
regard, we note a weak indication for two related modes
\cite{bergfeld}:

\begin{eqnarray}
{\cal B}(B^+\to \omega K^+) & = & (1.5^{+.7}_{-.6} \pm.2) \times 10^{-5} \\
{\cal B}(B\to\phi K^\ast) & = & (1.1^{+.6}_{-.5}\pm.2) \times 10^{-5}
\end{eqnarray}

\noindent in which the penguin contribution is expected to be dominant.
Based on these results we should also expect 

\begin{equation}
{\cal B}(B\to \omega
K^\ast) \simeq {\cal B}(B\to \rho K^\ast) \simeq 1\times10^{-5}
\end{equation}

\noindent which is also quite close to ${\cal B}(B\to K\pi)$ found by
CLEO \cite{cleoref}.

For $B^+\to \rho^+\omega$ we expect the tree graph to dominate. CLEO
finds \cite{jroy}

\begin{equation}
{\cal B}(B\to \pi^+\pi^0) = (1.5^{+.8}_{-.7} )\times 10^{-5} 
\end{equation}

\noindent Therefore it is quite likely that ${\cal B}(B\to \rho\omega)$
is also in the same ball-park.

On the other hand $B\to \rho\phi$ results from final state rescattering
effects. Most likely the branching ratio of this mode will,     
therefore, be smaller
than $10^{-5}$, by factors of order (3--10).

\item Our second comment deals with the application of our method for
extracting the CKM phase $\gamma$. Since in the standard phase
convention \cite{wolfmatrix} 
$\delta_T=\gamma$ and the $B_s$-$\bar B_s$ oscillation phase
$\hat \beta_{B_s\mbox{-}\bar B_s} =0$, it is therefore clear that to 
determine $\gamma$ through the use of our
$VV$ method will require the study of $B_s$
decays. As in the case of $B$ decays reported in this work, an
interplay of direct and mixing-induced CP (through $B_s$-$\bar B_s$
oscillations) will have to be involved. 
By inspection of the tree process $b\to
u\bar us$, which donates $\delta_T$, one immediately arrives at two
examples:

\begin{eqnarray}
(1) & & B_s\to K^\ast\rho \nonumber \\
(2) & & B_s\to \bar K^\ast K^\ast 
\end{eqnarray}

\noindent which can be used. We hope to return to these in a future
publication. 

\end{enumerate}

This research was supported in part by DOE contracts DE-AC02-98CH1- 0886
(BNL) and DE-FG02-94ER40817 (ISU).


\begin{thebibliography}{99}
 
\bibitem{cleoref} See, e.g., K. Lingel, T. Skwarnicki and J.G. Smith,
hep-ex/9804015, submitted to Ann.\ Rev.\ Nucl.\ and Part.\ Sci. 

\bibitem{article} See the article by H. Quinn, in Review of Particle
Properties, R.M. Barnett {\it et al}., Phys.\ Rev.\ {\bf D54}, 1
(1996). 

\bibitem{gronautwo} M. Gronau and D. London, Phys.\ Rev.\ Lett.\
{\bf65}, 3381 (1990). 



\bibitem{angles} 
Recall that, 
$\alpha \equiv \arg (-V_{td}V^\ast_{tb} / V_{ud}V^\ast_{ub})$,
$\beta \equiv \arg (-V_{cd}V^\ast_{cb}/ V_{td}V^\ast_{tb})$ and $\gamma
\equiv \arg(-V_{ud}V^\ast_{ub} / V_{cd}V^\ast_{cb})$; see e.g.\ Ref.~2.

\bibitem{fleischer} R. Fleischer, Z. Phys.\ {\bf C62}, 81 (1994); N.G.
Deshpande and X.-G. He, Phys.\ Rev.\ Lett.\ {\bf74}, 26 (1995); M.
Gronau, O.F. Hernandez, David London and J. Rosner, Phys.\ Rev.\ {\bf
D52}, 6374  (1995); A. J. Buras and R. Fleischer, Phys. Lett. 
{\bf B365}, 390 (1996).  

\bibitem{valen} For analysis of two vector final states see e.g.:  G. Valencia,
Phys.\ Rev.\ {\bf D39}, 3339 (1989); I. 
Dunietz, H.R. Quinn, A. Snyder, W. Toki and H. Lipkin,
Phys.\ Rev.\ {\bf D43}, 2193 (1991); R. Aleksan, I. Dunietz and B. Kayser, 
Z. Phys.\
{\bf C54}, 653 (1992) and G. Kramer, W.F. Palmer and H. Simma, Nucl.\
Phys.\ {\bf B428}, 77 (1994).  

%
%
\bibitem{wolfmatrix}
L.~Wolfenstein, Phys. Rev. Lett. {\bf 51}, 1945 (1984). 


\bibitem{pertphase} M.~Bander, D.~Silverman and A.~Soni, Phys.\ Rev.\
Lett.\ {\bf 43}, 242 (1979).  

\bibitem{donoghue} J.~F.~Donoghue, E.~Golowich, A.A.~Petrov and J.M.~Soares,  
Phys.\ Rev.\ Lett.\  {\bf 77}, 2178 (1996). 

\bibitem{mancp} D.~Atwood and A.~Soni, Phys.\ Rev.\ {\bf D58}, 036005 (1998).
For final state  rescattering
effects see also\cite{neubert}.   

\bibitem{neubert} J.-M. G\'erard and J. Weyers, hep-ph/9711469; M.
Neubert, Phys.\ Lett.\ {\bf B424}, 152 (1998); A. Falk 
A.L. Kagan, Y. Nir and A.A. Petrov, Phys.\ Rev.\ {\bf D57}, 4290 (1998)
and M. Gronau and J. Rosner, Phys.\ Rev.\ {\bf D57}, 2752 (1998).   

\bibitem{gronau} Y.~Nir and H.~Quinn, Phys. Rev. Lett. {\bf 67}, 541 (1991);
M.~Gronau, Phys. Lett. {\bf B265}, 389 (1991). 

\bibitem{bergfeld} T. Bergfeld {\it et al}. [CLEO Collaboration],
Phys.\ Rev.\ Lett.\ {\bf81}, 272 (1998). 

\bibitem{jroy} See the talks by J. Roy and J. Alexander at the ICHEP98,
Vancouver (July, 1998).  

\end{thebibliography}
\end{document}